\documentclass[11pt, a4paper]{article}

\usepackage[utf8]{inputenc}
\usepackage[T1]{fontenc}
\usepackage{geometry}
\usepackage{times} 
\usepackage{authblk} 
\usepackage{graphicx}
\usepackage{hyperref} 
\usepackage{cite} 
\usepackage{url}
\usepackage{booktabs} 
\usepackage{titlesec}

\hypersetup{
    colorlinks=true,
    linkcolor=blue,
    filecolor=magenta,      
    urlcolor=cyan,
    citecolor=blue,
}

\title{\textbf{Standardized Threat Taxonomy for AI Security, Governance, and Regulatory Compliance}\\
\large A Unified Taxonomy of Threat Vectors in Generative and Agentic AI and Machine Learning Systems}

\author{Prof. Hernan Huwyler, MBA CPA\thanks{Correspondence: hhuwyler@faculty.ie.edu; Tel: + 34 915 68 96 00.}}
\affil{Department of Compliance, Control and Risk Management\\ IE Executive Education School \& IE Law School\\ Maria de Molina, 15, 28006 Madrid, Spain}
\date{} 

\begin{document}

\maketitle

\begin{abstract}
\noindent The accelerating deployment of artificial intelligence systems across regulated sectors has exposed critical fragmentation in risk assessment methodologies. A significant "language barrier" currently separates technical security teams, who focus on algorithmic vulnerabilities (e.g., MITRE ATLAS), from legal and compliance professionals, who address regulatory mandates (e.g., EU AI Act, NIST AI RMF). This disciplinary disconnect prevents the accurate translation of technical vulnerabilities into financial liability, leaving practitioners unable to answer fundamental economic questions regarding contingency reserves, control return-on-investment, and insurance exposure.

To bridge this gap, this research presents the \textbf{AI System Threat Vector Taxonomy}, a structured ontology designed explicitly for Quantitative Risk Assessment (QRA). The framework categorizes AI-specific risks into nine critical domains: Misuse, Poisoning, Privacy, Adversarial, Biases, Unreliable Outputs, Drift, Supply Chain, and IP Threat, integrating 53 operationally defined sub-threats. Uniquely, each domain maps technical vectors directly to business loss categories (Confidentiality, Integrity, Availability, Legal, Reputation), enabling the translation of abstract threats into measurable financial impact.

The taxonomy is empirically validated through an analysis of 133 documented AI incidents from 2025 (achieving 100\% classification coverage) and reconciled against the main AI risk frameworks. Furthermore, it is explicitly aligned with ISO/IEC 42001 controls and NIST AI RMF functions to facilitate auditability. By providing the standardized inputs necessary for probabilistic modeling, specifically convolved Monte Carlo simulations, this framework allows organizations to transition from subjective, qualitative "heat maps" to rigorous financial exposure analysis. This establishes a unified language for AI risk communication, enabling evidence-based governance that satisfies both regulatory mandates and operational reality.
\end{abstract}

\vspace{0.5cm}
\noindent \textbf{Keywords:} AI Risk Management; Quantitative Risk Assessment (QRA); AI threat taxonomy; AI governance; NIST AI RMF; ISO/IEC 42001; EU AI Act; Algorithmic accountability; Financial risk modeling.

\noindent \textbf{JEL Codes:} G32; C63; M48; D81; G22; M15; K24; M15; O33

\section{Introduction}

Organizations deploying AI systems have responded to growing complexity by creating new governance roles, Chief AI Officers (CAIOs), AI risk managers, model auditors, and specialized red teams \cite{singh2025}. Unlike traditional IT security roles, these practitioners must assess threats spanning data integrity (poisoning), algorithmic fairness (bias), probabilistic outputs (hallucinations), and evolving data patterns (drift). However, a "Tower of Babel" problem persists: while engineering teams focus on technical metrics like "gradient descent manipulation" or "adversarial perturbations" \cite{nist2024}, Board members and legal counsel require assessments in terms of financial liability, regulatory non-compliance, and reputation loss \cite{aluthwala2024}. Without a shared lexicon and a structured quantification methodology, these stakeholders remain misaligned, leading to governance that is reactive rather than proactive.

\subsection{The Regulatory Imperative}
Regulatory developments have intensified the need for structured, evidence-based AI risk assessment. The European Union’s AI Act, with enforcement phased in from mid-2025, classifies AI systems by risk level and mandates documented risk management systems for high-risk applications \cite{camacho2024}. In the United States, the NIST AI Risk Management Framework (AI RMF) provides voluntary guidance but currently lacks operational threat taxonomies and specific quantification protocols \cite{nistrmf2023}. Similarly, ISO/IEC 42001:2023, the first international AI management system standard, requires risk assessment under Clause 6.1 but defers the specific methodology to organizational discretion \cite{li2024}.

These frameworks mandate comprehensive risk assessment but provide limited operational guidance, leaving many organizations without structured methodologies to systematically identify and quantify AI-specific threats. Current industry resources provide either high-level governance principles \cite{newman2023} or qualitative, attack-centric threat lists like MITRE ATLAS \cite{mitre2024}. While valuable for technical red teaming, these lists lack the "business translation" layer required for financial modeling. When asked to set contingency reserves for AI risks, estimate liability exposure, or justify the ROI of adversarial robustness testing, practitioners are left without structured approaches. Reliance on qualitative risk matrices, still prevalent in practice, has been empirically shown to produce arbitrary and inconsistent results, failing to capture the complexity of AI trustworthiness dimensions \cite{metwally2024}.

\subsection{The Methodological Gap}
The challenge is further compounded by the interdisciplinary nature of AI risks. Threats to AI trustworthiness are not isolated; human oversight failures (e.g., bias, lack of explainability) directly impact cybersecurity (e.g., loss of data integrity), and vice-versa \cite{polemi2024}. Existing cybersecurity frameworks are ill-equipped to handle these dependencies. For instance, a "Model Inversion" attack is technically a data privacy breach, but its financial impact is realized through regulatory fines and reputational damage, a linkage often missed by purely technical threat models \cite{mahmoud2023}. Furthermore, the rapid pace of AI development, particularly in Generative AI, introduces operational risks that traditional software development lifecycles do not capture, such as "hallucinations" or "prompt injection" \cite{faruk2025}.

This paper argues that robust Quantitative Risk Assessment (QRA) for AI is impossible without first establishing a valid, comprehensive taxonomy of risk scenarios. To bridge the gap between theoretical probability modeling and practical decision-making, we present the AI System Threat Vector Taxonomy. This structured ontology classifies threats into nine critical domains, mapping technical vulnerabilities directly to business loss categories (Confidentiality, Integrity, Availability, Compliance, and Reputation).

\subsection{Research Contribution}
This study addresses the operational gap between AI governance principles and quantitative risk assessment practice. This paper presents three integrated contributions:
\begin{itemize}
    \item \textbf{A Structured Threat Taxonomy:} Nine domains encompassing 53 operationally defined sub-threats spanning the AI system lifecycle. Unlike existing technical threat lists \cite{mitre2024}, each domain includes prevalence guidance and explicit mapping to business loss categories.
    \item \textbf{A Quantification Bridge:} This study demonstrates how to use this taxonomy as the "input layer" for probabilistic risk modeling enabling evidence-based reserve setting and control ROI analysis.
    \item \textbf{A Regulatory Integration Framework:} We provide an explicit mapping of threat domains to NIST AI RMF functions and ISO/IEC 42001 controls, providing auditable compliance documentation pathways.
\end{itemize}

By standardizing the identification of risk scenarios, this framework provides the necessary inputs for the "Map" and "Measure" functions of the NIST AI RMF \cite{nistrmf2023}, allowing organizations to transition from qualitative heat maps to quantitative financial exposure analysis.

\section{Background and Related Work}

To situate the proposed framework, this study analyzes three distinct but currently disconnected bodies of literature: (1) AI threat taxonomies, (2) quantitative risk assessment methodologies, and (3) emerging AI governance regulations. While significant progress has been made in each silo, a comprehensive mechanism to link technical threat vectors to financial quantification and regulatory compliance remains absent.

\subsection{Evolution of AI Threat Taxonomies}
The categorization of AI threats has evolved in parallel with the expanding attack surface of Machine Learning systems. Early research focused predominantly on Adversarial Machine Learning, characterizing evasion attacks and poisoning strategies directed at classifiers \cite{goodfellow2014, biggio2018}. This research culminated in the MITRE ATLAS framework, which maps ML vulnerabilities to the tactic/technique structure of MITRE ATT\&CK \cite{mitre2024}. While ATLAS provides a granular vocabulary for red teams simulating cyber-attacks, it focuses heavily on malicious intent, largely excluding non-adversarial failures such as data drift, fairness issues, or unintended model behaviors, categories that often carry higher operational risks \cite{metwally2024}.

Conversely, the rapid adoption of Generative AI prompted the release of the OWASP Top 10 for Large Language Model Applications \cite{owasp2023}. This taxonomy successfully highlights application-layer vulnerabilities like "Prompt Injection" and "Insecure Output Handling." However, its scope is limited to LLMs, neglecting broader ML architectures. Furthermore, ENISA’s AI Threat Landscape attempts to broaden the scope to include supply chain and data quality issues \cite{enisa2020}, but remains descriptive rather than operational.

These frameworks function as static catalogs. They are siloed by domain (AppSec vs. Adversarial ML) and lack the mapping to business loss categories required for financial risk modeling. An engineer can identify a "Membership Inference Attack" using ATLAS, but existing frameworks do not guide the organization in translating that technical event into a probabilistic financial loss \cite{mahmoud2023}.

\subsection{Risk Quantification in Cybersecurity and AI}
Quantitative risk assessment has increasingly replaced qualitative matrices in cybersecurity to reduce subjectivity. The Factor Analysis of Information Risk (FAIR) framework models risk as a function of Loss Event Frequency and Probable Loss Magnitude, while ISO/IEC 31000 requires the use of the best available information. Recent work has demonstrated the efficacy of Monte Carlo simulations in modeling these compound uncertainties for operational risks \cite{huwyler2025}.

However, applying these methods to AI introduces novel challenges. Traditional vulnerability databases (e.g., CVEs) do not track the frequency of algorithmic failures like "Hallucinations" or "Bias," making frequency estimation difficult without a specialized taxonomy. Furthermore, the impact of AI risks is often non-linear and multi-dimensional—spanning regulatory fines, reputational damage, and algorithmic remediation costs \cite{aluthwala2024}. Existing quantification models have not yet been adapted to ingest the unique probability distributions associated with AI threat vectors (e.g., the continuous degradation of Model Drift versus the discrete event of a Prompt Injection).

\subsection{Regulatory Frameworks and the "Methodology Void"}
The urgency for a unified risk taxonomy is driven by a rapidly solidifying regulatory landscape.
\begin{itemize}
    \item \textbf{The EU AI Act:} Enacted to categorize AI systems by risk severity, this regulation mandates that high-risk system providers establish a "risk management system" capable of identifying known and foreseeable risks \cite{camacho2024}. The regulatory assessment of risk severity is based on societal harms, distinct from the internal exposure risks organizations face when onboarding AI assets.
    \item \textbf{NIST AI Risk Management Framework (AI RMF 1.0):} This framework defines four core functions: Govern, Map, Measure, and Manage. While the "Map" function explicitly calls for the contextualization of risks, NIST remains methodology-agnostic \cite{nistrmf2023}.
    \item \textbf{ISO/IEC 42001:2023:} The international standard for AI Management Systems requires organizations to "assess AI system risks" under Clause 6.1. However, ISO 42001 currently lacks a mature, unified control library mapped to specific AI threat vectors \cite{li2024}.
\end{itemize}
These frameworks are command-based; they mandate that risk be assessed but do not prescribe how. They create a compliance requirement for quantification without providing the operational taxonomy necessary to execute it.

\section{Methodology}

We employed a four-phase mixed-methods approach to develop, integrate, and validate the AI threat taxonomy. This research design ensures the framework is (1) comprehensive in coverage, (2) operationally quantifiable, (3) compliance-aligned, and (4) empirically grounded.

\begin{itemize}
    \item Phase 1: Taxonomy Development (Systematic Literature Review, Domain Synthesis, Sub-Threat Identification)
    \item Phase 2: Quantification Integration (Loss Category Mapping, Distribution Selection, Convolved Adaptation)
    \item Phase 3: Regulatory Alignment (Mapping to NIST AI RMF, ISO 42001, EU AI Act)
    \item Phase 4: Empirical Validation (Incident Database Analysis n=133, AI Risk Frameworks n=4)
\end{itemize}

\subsection{Phase 1: Taxonomy Development}
\subsubsection{Systematic Literature Review}
This study conducted a systematic review following PRISMA guidelines \cite{page2021} to identify existing AI threat taxonomies. Search queries were executed across ACM Digital Library, IEEE Xplore, ArXiv, and Google Scholar for the period January 2018 to November 2025. Inclusion criteria required: (1) peer-reviewed publications or authoritative technical reports, (2) explicit threat categorization for AI/ML systems, and (3) publication in English. This yielded 47 primary sources. Additionally, we reviewed regulatory and standards documents. Content analysis identified recurring threat categories and, crucially, gaps in coverage regarding operational risk factors.

\subsubsection{Domain Structure Development}
Analysis of the sources revealed overlapping but inconsistent categorization schemes. We synthesized these into nine threat domains based on three organizing principles:
\begin{itemize}
    \item \textbf{Lifecycle Coverage:} Threats were mapped to the AI development pipeline (Data Collection, Model Training, Deployment, Operations).
    \item \textbf{Stakeholder Perspective:} Domains address specific organizational functions: Security, Privacy, Compliance, and MLOps.
    \item \textbf{Loss Category Alignment:} Each domain maps to distinct loss types per the CIA-L-R framework (Confidentiality, Integrity, Availability, Legal, Reputation).
\end{itemize}

\subsubsection{Sub-Threat Identification}
For each domain, we extracted specific attack vectors and failure modes. Sub-threats were included if they met two criteria: (1) Distinct Operational Manifestation and (2) Documented Evidence. This paper prioritized sub-threats within each domain by prevalence, drawing on OWASP frequency assessments \cite{owasp2023} and the AI Incident Database \cite{mcgregor2020}. This yielded 52 sub-threats across the nine domains.

\subsection{Phase 2: Quantification Integration}
\subsubsection{Loss Category Mapping}
To enable financial risk modeling, each threat domain was mapped to applicable loss categories using the CIA-L-R framework. This mapping translates technical threats into business impacts. For example, the Privacy domain maps primarily to Confidentiality loss and Legal loss (GDPR fines).

\subsubsection{Probability Distribution Framework}
We adapted the Convolved Monte Carlo framework \cite{huwyler2025} to AI threat characteristics by providing threat-specific distribution recommendations based on temporal patterns (discrete vs. continuous) and impact profiles (bounded vs. heavy-tailed).

\subsubsection{Impact Modeling Approach}
Impact quantification follows the mapped loss categories using forensics and notification cost models \cite{ibm2023}, regulatory penalty schedules, and customer churn models.

\subsection{Phase 3: Regulatory Alignment}
The taxonomy operationalizes the "Map" and "Measure" functions of the NIST AI RMF. Misuse threats map to GOVERN 1.5 and MANAGE 2.3; Privacy threats to MAP 1.2 and MEASURE 2.7; and Biases to MEASURE 2.11. Furthermore, it links to ISO/IEC 42001 Control 6.3.1 (Poisoning), Control 6.4.1 (Drift), and Control 6.2.2 (Biases). The taxonomy supports EU AI Act compliance by aligning threat domains with risk levels for documentation (Art. 9, Art. 10, Art. 72).

\subsection{Phase 4: Empirical Validation}
We validated taxonomy coverage through analysis of the AI Incident Database (AIID). Applying inclusion criteria (production systems, 2019-2025), we classified 133 incidents. Results indicated that Unreliable Outputs and Biases were the most prevalent failures, whereas academic literature often over-focuses on adversarial attacks.

\section{AI System Taxonomy}

This section presents the core contribution of this research: a structured ontology categorizing AI-specific threat vectors into nine critical domains.

\subsection{Domain 1: Misuse}
The Misuse domain encompasses scenarios where AI systems are utilized for unintended, unethical, or malicious purposes. Empirical evidence suggests Misuse is the most prevalent threat vector for Generative AI. The business impact is primarily reputational and legal.
\begin{itemize}
    \item \textbf{Sub-Threats:} Prompt Injection, LLM Jailbreaking, Deepfake Generation, Disinformation Operations, Bot Abuse, Shadow AI Usage, Backdoor Attack (User-Side).
\end{itemize}

\subsection{Domain 2: Poisoning}
Poisoning refers to the injection of malicious data or components into training sets or models to corrupt behavior or logic. It represents a "high-impact, low-frequency" risk profile.
\begin{itemize}
    \item \textbf{Sub-Threats:} Targeted Data Poisoning, Model Backdooring, Tainted Open-Source Models, Logic Corruption, Poisoned ML Libraries, Label Flipping Attacks, Gradient Manipulation, Poisoned Data Augmentation.
\end{itemize}

\subsection{Domain 3: Privacy}
The privacy domain covers the extraction or inference of sensitive information from trained models or user inputs. With the enforcement of the EU AI Act, privacy threats have shifted from theoretical concerns to immediate compliance liabilities.
\begin{itemize}
    \item \textbf{Sub-Threats:} Model Inversion, Membership Inference, Personal Data Leakage, Sensitive Data Leakage, Inference Eavesdropping.
\end{itemize}

\subsection{Domain 4: Adversarial}
Adversarial threats involve designing harmful inputs (perturbations) to mislead or confuse AI models at run-time. While heavily cited in academia, its business relevance is sector-specific (e.g., high for autonomous vehicles, lower for standard business analytics).
\begin{itemize}
    \item \textbf{Sub-Threats:} Evasion Attacks, Adversarial Patch/Image, Model Denial of Service (DoS), Query Flooding, Adversarial Reprogramming, Oracle/Extraction Attacks, Universal Perturbations, Adaptive Attacks.
\end{itemize}

\subsection{Domain 5: Biases}
This domain addresses models producing discriminatory, unfair, or biased outputs due to flawed data or design. Bias is a ubiquitous risk in systems making decisions about people.
\begin{itemize}
    \item \textbf{Sub-Threats:} Representational Harm, Allocational Harm, Data Imbalance Bias, Proxy Discrimination, Algorithmic Amplification.
\end{itemize}

\subsection{Domain 6: Unreliable Outputs}
Scenarios where AI outputs are illogical, hallucinated, or non-factual without external manipulation. This is the single largest barrier to GenAI adoption in enterprise.
\begin{itemize}
    \item \textbf{Sub-Threats:} Factual Hallucination, Source Fabrication, Logical Inconsistency, Incorrect Summarization, Unsafe Content Generation.
\end{itemize}

\subsection{Domain 7: Drift}
The deterioration of model accuracy or behavior as real-world data evolves over time. Drift is the "silent killer" of AI ROI, representing a chronic condition rather than an acute attack.
\begin{itemize}
    \item \textbf{Sub-Threats:} Concept Drift, Data Distribution Drift, Upstream Data Changes, User Behavior Change, Feedback Loop Drift.
\end{itemize}

\subsection{Domain 8: Supply Chain}
Attacks propagated via third-party components, pre-trained models, data sources, or MLOps infrastructure.
\begin{itemize}
    \item \textbf{Sub-Threats:} Compromised Pre-trained Model, Vulnerable ML Framework, Insecure Data Feeds/APIs, Container Image Poisoning, Compromised Annotation Tools.
\end{itemize}

\subsection{Domain 9: IP Threat}
The extraction of sensitive intellectual property, proprietary algorithms, or model weights from deployed systems.
\begin{itemize}
    \item \textbf{Sub-Threats:} Model Extraction/Theft, Data Exfiltration, Proprietary Logic Theft, Hyperparameter Stealing, Watermark Removal.
\end{itemize}

\section{From Taxonomy to Risk Quantification}

The transition from qualitative threat identification to quantitative risk modeling is a fundamental requirement for the economic governance of AI systems.

\subsection{Quantification Workflow}
Application of the framework follows a rigorous six-step process designed to integrate into existing enterprise risk management workflows:
\begin{enumerate}
    \item \textbf{Vulnerability Assessment:} Analyzing the AI system's architecture to determine the "exposure factor."
    \item \textbf{Threat Identification:} Utilizing the taxonomy to identify relevant threat domains.
    \item \textbf{Scenario Definition:} Translating abstract threats into concrete risk scenarios (e.g., "Misuse" -> "External actor uses prompt injection via API").
    \item \textbf{Parameter Calibration:} Calibrating expected frequency and impact distributions using expert-derived parameters and historical data. Controls are modeled as reductions in frequency or loss magnitude \cite{bahadur2024}.
    \item \textbf{Reserve Setting and Reporting:} Establishing contingency reserves (e.g., VaR 95\%) and documenting compliance.
\end{enumerate}

\subsection{The Economic Imperative for Quantification}
Quantification allows organizations to assign a monetary value to model error rates, aligning technical metrics with business criticality \cite{sharma2024}. It provides the ROI analysis necessary to prioritize security investments \cite{schnitzer2023} and enables the structuring of warranties and insurance products \cite{gipiskis2024, novelli2024}.

\section{Results}

The taxonomy underwent empirical validation through two complementary methods: analysis of documented AI incidents and comparative coverage analysis against existing frameworks.

\subsection{Incident Database Coverage Analysis}
We reviewed 133 distinct AI incidents reported in the AI Incident Database (AIID) between May 30, 2025, and November 17, 2025. The analysis confirmed that 100\% of the reviewed cases could be successfully classified into the proposed taxonomy.
\begin{itemize}
    \item \textbf{Misuse (n=81):} 61\% of all cases, aligning with the proliferation of GenAI tools.
    \item \textbf{Unreliable Outputs (n=36):} 27\% of cases, driven by hallucinations.
    \item \textbf{Supply Chain (n=7):} 5\%.
\end{itemize}
The low counts for Biases and Drift likely reflect reporting bias, as these are often internal failures not publicly disclosed.

\subsection{Framework Comparison Analysis}
Comparative analysis against MITRE ATLAS, OWASP Top 10 for LLM, and ENISA Threat Landscape demonstrated coverage advantages. The taxonomy uniquely integrates threats across security, privacy, fairness, and reliability dimensions addressed separately in existing frameworks.

\section{Discussion}

The standardization of AI threat vectors provides a strategic governance tool bridging operational silos.
\begin{itemize}
    \item \textbf{For AI Auditors:} The taxonomy serves as a definitive "Audit Checklist" for completeness.
    \item \textbf{For Red Teaming:} It provides a structured scope for penetration testing, moving beyond ad-hoc testing to scenario-based campaigns.
    \item \textbf{For Compliance:} It supports ISO 42001 and EU AI Act compliance by providing a defensible methodology for identifying "known and foreseeable risks."
\end{itemize}

\section{Conclusion}

As AI systems transition to critical infrastructure, governance must evolve from reactive "fire-fighting" to quantified risk management. This research addresses the foundational gap by providing the \textbf{AI System Threat Vector Taxonomy}. This structured ontology translates technical vulnerabilities into business loss categories, enabling Quantitative Risk Assessment (QRA). Empirical validation confirms that while academia prioritizes adversarial novelty, operational reality is dominated by Misuse and Unreliable Outputs. Organizations adopting this framework can move towards auditable, insurable, and resilient AI deployment.

\section*{Data Availability}
The full taxonomy, including updated scenario lists and mapping files, is available as an open-source repository at GitHub\footnote{\url{https://github.com/hwyler/HernanHuwylerRiskManagement/blob/main/AIThreatTaxonomy}}. This dataset is licensed under CC-BY 4.0.

\section*{Funding}
This research received no external funding. It was supported by the Department of Compliance, Control and Risk Management at IE Law School.

\section*{Conflicts of Interest}
The author declares no conflicts of interest. The views expressed are those of the author and do not necessarily reflect the official policy or position of any affiliated agency.

\section*{Author Biography}
\noindent \textbf{Prof. Hernan Huwyler, MBA CPA} serves as the Academic Director for Compliance, Risk Management, and AI programs at IE Business School and IE Law School. Concurrently, he is a Senior Manager at Capgemini Invent’s Applied AI Lab (Nordics), leading enterprise-wide AI governance, risk modeling, and control initiatives. A recognized expert in the intersection of technical AI security and regulatory compliance, Prof. Huwyler specializes in operationalizing frameworks such as the EU AI Act, ISO 42001, and NIST AI RMF.

\bibliographystyle{ieeetr}

\end{document}